\def\year{2023}\relax
\documentclass[letterpaper]{article} 
\usepackage{booktabs}
\usepackage{multirow}
\usepackage[noend]{algpseudocode}
\usepackage{cite}
\usepackage{balance}
\usepackage{algorithmicx,algorithm}
\renewcommand{\algorithmicrequire}{\textbf{Input:}}  
\renewcommand{\algorithmicensure}{\textbf{Output:}} 
\usepackage{aaai23}  
\usepackage{times}  
\usepackage{helvet} 
\usepackage{courier}  
\usepackage[hyphens]{url}  
\usepackage{graphicx} 
\usepackage{caption} 
\title{DCRNN: A Deep Cross approach based on RNN \\ for Partial Parameter Sharing in Multi-task Learning}
\author{
    Jie Zhou,\textsuperscript{\rm 1, 2}
    Qian Yu,\textsuperscript{\rm 2}\equalcontrib
}

\affiliations{
    \textsuperscript{\rm 1} Xiaomi Inc, China\\
    \textsuperscript{\rm 2} School of Software, Beihang University, China\\

\{zhoujiee, qianyu\}@buaa.edu.cn
}

\begin{document}
	
\maketitle
	
	\begin{abstract}
		In recent years, deep learning has developed rapidly, and personalized services are also actively exploring using deep learning algorithms to improve the performance of the recommendation system. For personalized services, a successful recommendation consists of two parts: Attracting user to click the item, and the user is willing to consume the item (content consumption, product ordering, etc.). If both tasks need to be predicted at the same time, traditional recommendation system generally trains two independent models, one is to predict the user's click behavior, the other is to predict the user's consume behavior. This approach is cumbersome and does not effectively model the relationship between the two subtasks of "click-consumption". Therefore, in order to improve the success rate of recommendation and reduce the computational cost, researchers are trying to model multi-task learning. \\
		At present, existing multi-task learning models generally adopt hard parameter sharing or soft parameter sharing architecture. But these two architectures each have certain problems. Hard parameter sharing cannot express each target well, and there is a risk of underfitting. While in the soft parameter sharing model, the size of the network will increase linearly as the number of tasks increases, and the requirements for computing resources will also increase. In addition, since many features in the recommendation system are calculated or designed by human beings, some noises will inevitably be introduced, which will weaken the learning ability of the model. So, in this work, we propose a novel recommendation model based on real recommendation scenarios, \textit{Deep Cross network based on RNN for partial parameter sharing} (DCRNN). The model has three innovations: 1. It adopts the idea of cross network, uses RNN network to cross-process the features, thereby effectively improves the expressive ability of the model; 2. It innovatively proposes the structure of partial parameter sharing, which effectively combines the advantages of hardware and software parameter sharing; 3. By using feature adaptation function, the model can adaptively learn original features, which enhances the model’s anti-noise ability. We experimented with the proposed model and the baseline model on the real scenario of Xiaomi Radio’s recommendation and the Ali-CCP dataset. The results show that DCRNN outperforms the baseline model. \\
		
	\end{abstract}
	
	\section{Introduction}
	Recommendation system is an effective tool to help users solve the problem of information overload. It has been widely applied in many commercial fields, including advertising computing, social network, e-commerce \cite{wen2019multi,covington2016deep}. The recommendation algorithm is the core of the system, which determines the performance of the recommendation system. \\
	The traditional recommendation algorithm takes click-through rate (CTR) as the optimization target, which helps users to find the most favorite commodity or the most valuable information. Such as logistic regression, decision tree algorithm \cite{chen2015xgboost,ke2017lightgbm}, factorization machine \cite{rendle2010factorization,juan2016field} in machine learning, and wide\&deep model \cite{cheng2016wide}, DeepFM \cite{guo2017deepfm}, and deep interest network (DIN) \cite{zhou2018deep} in deep learning. \\
	However, in personalized services, a successful recommendation consists of two subtasks: attract users to click on the recommended item; the recommended item meets the user's expectations and is consumed by the user. The general approach is to train two independent models, one is to predict whether the user clicks (CTR), and the other is to predict whether the user consumes (CVR) \cite{wen2020entire}. However, this approach is too cumbersome and does not model the correlation between the two tasks. Therefore, in order to improve the accuracy of the task and reduce the computational cost, researchers have begun to explore using deep learning to build a multi-task learning model.\\
	\begin{figure*}[t]
		\centering
		\includegraphics[scale=0.4]{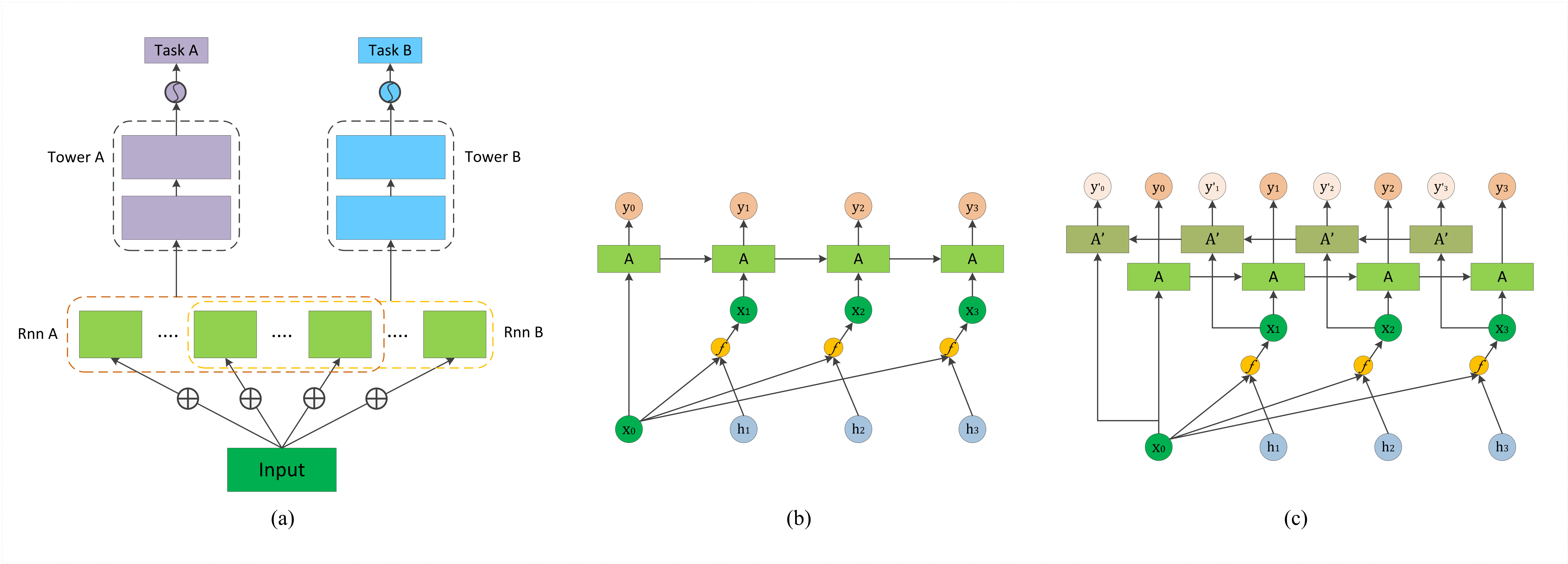} 
		\caption{(a) DCRNN model architecture; (b) CRNN(Unidirectional); (c)CRNN(Bidirectional)}
		\label{fig1}
	\end{figure*}
	At present, many multi task learning models have been proposed. For example, Jiaqi Ma et al. \cite{ma2018modeling} proposed Multi-gate Mixture-of-Experts (MMoE), which has attracted great attention in the industry. In MMOE, each task shares the same expert layer in the bottom layer. In the upper layer, the gating network is used to control the weight of expert layer, so as to achieve the purpose of modeling different tasks. In many real recommendation scenarios, such as e-commerce scenarios, users' click behavior and purchase behavior have strong sequence correlation \cite{wen2020entire,ma2018entire}. However, MMOE model has obvious advantages in multi task learning with low correlation tasks, and when the correlation between tasks is high, the advantage is weakened. In addition, although the MMOE adopts gating network, it is still a hard parameter sharing architecture, that is, each task shares a common experts layer, which limits the ability of the model to express the specific information of each task. \\
	In this paper, we propose a novel multi-task learning model, Deep Cross network based on RNN for partial parameter sharing (DCRNN). As shown in Fig. 1-(a), at the bottom layer of DCRNN, we use the input features to construct an input sequence as the input of RNNs in DCRNN, and we use RNN network to cross process the features. DCRNN achieves the goal of "partial parameter sharing" by modeling RNN layer separately for each task, and the subsequences of RNNs are partially overlapped, which implicitly learns the correlation of between tasks. Finally, the output of RNN will be used as the input of the tower of each task in the upper layer. In addition, we propose “feature adaptive function”. When constructing the input features sequence, we use the adaptive function to process the features in the sequence. The purpose is to enhance the anti-noise ability of the model by learning adaptive parameters, and make the features in the sequence better expressed when the model is trained. \\
	In order to test the performance of DCRNN in recommendation scenarios, we done experiments in Xiaomi Radio’s recommendation, and take MMOE as the baseline model. Experiments show that DCRNN outperforms the baseline model in both offline evaluation and online evaluation, and the number of parameters of DCRNN is far less than the baseline model, which is easy to deploy the model online. In addition, we also conducted experiments in a public “Ali-CCP” dataset\footnote{https://tianchi.aliyun.com/dataset/dataDetail?dataId=408}, and the experimental results shown that DCRNN outperforms the baseline model. \\
	The rest construction of this paper is as follows: Section 2 reviews the related work; Section 3 introduces the proposed DCRNN; Section 4 verifies the effectiveness of the proposed model in the real business of Xiaomi radio and the Ali-CCP dataset; Section 5 summarizes our work. \\

	\section{Related Work}
	\subsection{Feature Crossover}
	In previous work, Ruoxi Wang et al. \cite{wang2017deep} proposed \textit{Deep \& Cross Network} (DCN), which aims to explicitly model high-order feature crossovers. The feature interaction operations are as follows: \\
	\begin{equation}\label{dcn}H_{t}=X_{0} \cdot H_{t-1}^{T} \cdot W_{t-1}+b_{t-1}+H_{t-1}\end{equation}
	Where, $H_{t-1}$  and $H_{t}$ represent the $(t-1)-t h$ and $t-t h$ cross-hidden layers respectively, $W_{t-1}$ and $b_{t-1}$ are the weight matrix and offset parameters of the $(t-1)-t h$ cross-hidden layer. The iteration of Formula (\ref{dcn}) constitutes the main part of the deep cross network model. The above formula constitutes the main part of DCN. \\
	DCN can learn feature interaction well, however it can be seen in Formula (\ref{dcn}): there is a scalar multiple relationship between $H_{t}$ and $H_{t-1}$, which limits the learning ability of the DCN. Therefore, Jianxun Lian et al. \cite{lian2018xdeepfm} proposed "\textit{Compressed Interaction Network} (CIN)" on the basis of DCN. They thought that the interaction of features are applied at the vector-wise level, not at the bit-wise level, that is, features should be crossed on features field \cite{juan2016field}. The specific feature cross formula is as follows: \\
	\begin{equation}\label{cin}H_{t}=\sum_{i=1}^{H_{t-1}} \sum_{j=1}^{m} W_{i, j}\left(H_{t-1} \circ X_{0}\right)\end{equation}
	Where, $H_{t}$ is the cross-hidden layer, $W$ is the weight matrix, $X_{0}$ is the original feature, $\circ$ represents the Hardman product. \\
	Although the above two classic cross networks can effectively perform feature cross operations, they also have some common shortcomings: 
	\begin{itemize}
		\item  Large-scale parameters. In DCN and CIN, with the deepening of the hidden layer and the increase of the number of features, the parameters of the model will increase rapidly, which will lead to the model cannot be well deployed online, and it is difficult to serve personalized business. 
		\item The noise of the original feature can affect the expression ability of the model. It can be seen in Formula (\ref{dcn}) and (\ref{cin}) that the generation of cross layer is the same as the original features $X_{0}$. However, the feature processing will inevitably bring some noise to feature engineering. And the noise is continuously transmitted to the hidden layer in the iterative process, which will influense the ability of model learning.
	\end{itemize}
	\subsection{Modeling User Sequence by RNN}
	RNN has been proved to be very good at modeling sequence data and has been successfully applied to speech recognition \cite{miao2015eesen}, natural language processing \cite{yin2017comparative} and image processing \cite{shi2016end}. \\
	In some recommendation algorithms, RNN can well model the user's sequence behavior. Feng Yu et al. \cite{yu2016dynamic} successfully applied RNN to "basket recommendation". Jiaxi Tang et al. \cite{tang2019towards} used RNN and attention mechanism to model the dependence of long-term and long-term sequence behavior of users. However, different from the previous work on recommendation algorithm, we use RNN to model the shared feature sequences among multiple tasks, which is based on the cognition that "click-consume has certain sequence correlation". \\
	\subsection{Parameter sharing in multi-task learning}
	In deep learning, multi-task learning tries to model the common information among tasks in the hope that a joint task learning could result in better generalization performance. Therefore, unlike single task, which uses its own network alone to solve problem, multi-task learning is designed to learn shared parameters to optimize respective tasks. This has several advantages\cite{vandenhende2020revisiting}: 
	\begin{itemize}
		\item The shared learning mode of common network layer can reduce the use of computing memory.
		\item As they explicitly avoid to repeatedly calculate the features in the shared layers, once for every task, they show increased inference speeds.
		\item The mutual complementation and restriction of shared layers may improve the performance of the model.
	\end{itemize}
	In the previous work, parameter sharing in multi-task learning is divided into hard parameter sharing and soft parameter sharing. In hard parameter sharing, all tasks share the same sharing layer at the bottom, and each task has its own special operation in the upper layer. And in soft parameter sharing \cite{misra2016cross,liu2019end,gao2019nddr}, each task has its own separate parameter set (except for special operations), and there are some cross operations between these parameter sets. \\
	However, these parameter sharing seem to be two extreme way. Due to the fixed sharing layer, hard parameter sharing can not express each target well, which is prone to the risk of under fitting \cite{ruder2017overview}. Each task has its own independent parameter set, so soft parameter sharing may lead to the problem of model scalability, and the reason is that the scale of the multi-task learning network tends to increase linearly with the number of tasks.
	
	\begin{figure}[t]
		\centering
		\includegraphics[width=1.0\columnwidth]{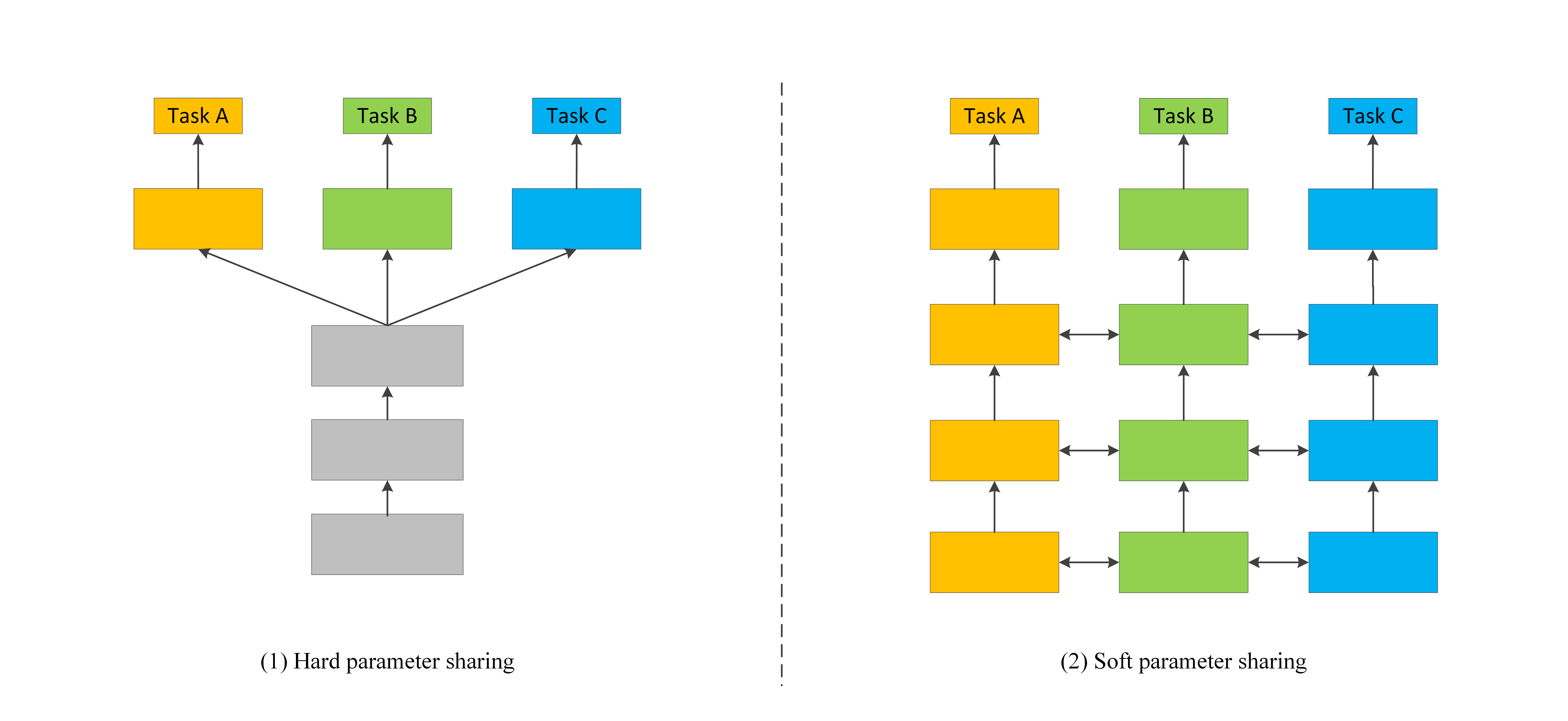}
		\caption{parameter sharing in multi-task learning}
		\label{fig2}
	\end{figure}
	
	\section{Modeling Approach}
	\subsection{Modeling Cross network by RNN}
	In Section 2, we introduce the feature cross, we can further summarize the cross network. In DCN, the construction of cross-hidden layers can be summarized as follows:\\
	\begin{equation}H_{t}=X_{0} \cdot H_{t-1}^{T} \cdot W_{h}+b+H_{t-1}=f\left(X_{0}, H_{t-1}\right)\end{equation}Similarly, in xDeepFM, the construction of CIN can be summarized as follows:\\
	\begin{equation}H_{t}=\sum_{i=1}^{H_{t-1}} \sum_{j=1}^{m} W_{i, j}\left(H_{t-1} \cdot X_{0}\right)=f\left(X_{0}, H_{t-1}\right)\end{equation}
	By observing the construction of the DCN and CIN, we can find that \textbf{the hidden layer $H_{t}$ of two models is a function about $X_{0}$ and $H_{t-1}$}. The construction of hidden layer in RNN is as follows:
	\begin{equation}\label{original_rnn}H_{t}=\varphi\left(X_{t} \cdot W_{t}+H_{t-1} \cdot W_{h}+b\right)\end{equation}
	Where, $\varphi$ is the activation function, $X_{t}$ is the input of the current time $t$, $H_{t-1}$ is the hidden layer output of the previous time $t-1$, $W_{i}$ is the weight matrix and b is a offset.\\
	Assume we make the input $X_{t}$ at each time in Formula (\ref{original_rnn}) becomes $X_{0}$, the above formula becomes:\\
	\begin{equation}H_{t}=\varphi\left(X_{0} \cdot W_{t}+H_{t-1} \cdot W_{h}+b\right)=f\left(X_{0}, H_{t-1}\right)\end{equation}
	Ingeniously, the hidden layer $H_{t}$ in RNN is also a function about $X_{0}$ and $H_{t-1}$. Therefore, we can conclude that \textbf{when $X_{t}=X{_0}$, the calculation process of RNN plays a similar role as DCN or CIN}. Therefore, we can apply RNN to feature cross processing, which is "Cross network based on RNN (CRNN)"(as shown in Figure 1-(b,c)).\\
	In the experimental part, we set RNN to LSTM / GRU and BiLSTM / BiGRU respectively for offline training.
	\subsection{Partial parameter sharing}
	Based on the problem of hard and soft parameter sharing described in Section 2, in order to find a compromise solution between hard and soft parameter sharing, we propose the concept of "partial parameter share" (It will be introduced in the next subsection). The specific method is as follows: we take the feature sequence as the input of RNN for each task, and each RNN can only obtain the input of partial feature sequence (subsequence). There are overlapping feature sequences among RNNs, and the overlapped part of feature sequences is shared (as shown in Figure Figure 1-(a)). \\
	
	\begin{algorithm}[ht]
		\caption{partial parameter sharing} 
		\hspace*{0.02in} {\bf Input:} 
		$n$: The number of tasks, $L$: The sequence length required for each RNN, $I$: Sequence intervals between RNNs\\
		\hspace*{0.02in} {\bf Input:} 
		$Seq$: features sequence\\
		\hspace*{0.02in} {\bf Output:} 
		$K$: Input features subsequences for RNNs 
		\begin{algorithmic}[1]
			\\Let $0 \leq I \leq L$       
			\\For {$i = 0$ to n do} 
			\\\quad $K_{i}={Seq}[i * I: L+i * I]$
			\\END For\\
			\Return $K$
		\end{algorithmic}
		\label{algorithm_1}
	\end{algorithm}
	Algorithm \ref{algorithm_1} is the implementation of "partial parameter sharing". From the algorithm, we can see that soft parameter sharing and hard parameter sharing are special cases of algorithm \ref{algorithm_1}. When $I=0$, algorithm \ref{algorithm_1} is hard parameter sharing; when $I=L$, algorithm \ref{algorithm_1} is soft parameter sharing.
	\subsection{Feature adaptive processing}
	Feature engineering has attracted more and more attention because of its ability to improve the performance of various machine learning models. However, if the feature engineering is not designed properly, some noises will be introduced artificially, which will affect the training of the model and lead to the decline of the generalization ability of the model \cite{wang2017randomized}. Therefore, in order to improve the anti-noise ability and robustness of the model, we propose the concept of feature adaptive function in the process of constructing feature sequence, that is, we use the feature adaptive function to process the generated feature sequence. In this way, the expression ability and the anti-noise ability of the model can be enhanced by learning the adaptive parameters during the model training.\\
	We use $X_t$ represents the input of the new RNN at time $t$ after processing by the feature adaptive function $F(.)$. Feature adaptive function for original input $X_0$ The final formula of DCRNN hidden layer feature cross processing is as follows: \\
	\begin{equation}
	\left\{
	\begin{array}{ll}
	X_{t}=F\left(X_{0} ; a_{t}\right), S e q=\left[X_{0}, X_{1} \cdots X_{n}\right]\\
	H_{t}=\varphi\left(X_{t} \cdot W_{t}+H_{t-1} \cdot W_{h}+b\right)
	\end{array}
	\right.
	\end{equation}
	Where, $a$ is the feature adaptive parameter, and $Seq$ is the feature sequence constructed by the feature adaptive function. For the adaptive function $F(.)$, we use simple addition to process the original features.\\
	\begin{algorithm}[ht]
		\caption{Feature adaptive processing} 
		\hspace*{0.02in} {\bf Input:} 
		$X_{0}$: Original feature input, $n$: the length of $Seq$\\
		\hspace*{0.02in} {\bf Output:} 
		$Seq$: features sequence
		\begin{algorithmic}[1]
			\\For {$i = 0$ to n do}          
			\\\quad Create an adaptive parameter $A_{i}$
			\\\quad $X_{i}=F_{a d d}\left(X_{0} ; A_{i}\right)=X_{0} \oplus A_{i}$
			\\\quad $Seq$.append($X_{i}$)
			\\END For\\
			\Return $Seq$
		\end{algorithmic}
		\label{algorithm_2}
	\end{algorithm}
	
	\section{Experiments}
	\subsection{Xiaomi Radio’s recommendation}
	In this subsection, we report and discuss the experimental results of Xiaomi Radio’s recommendation.\\
	\subsubsection{Task Introduction}
	In order to test the performance of DCRNN, we conducted experiments on the "Album Recommendation" channel of the Xiaomi Radio APP. This scenario aims to recommend favorite albums for users and increase the playing time of users on this channel. From recommendation system, this scenario contains two tasks, click (album) and valid play (album). We constructed a DCRNN on click task and valid play task.\\
	\subsubsection{Data description}
	We used a real, large-scale dataset with 320 million samples, including 11 million users and 140, 000 albums. We randomly split the dataset into training set and test set with a ratio of 7:3.\\
	\subsubsection{Baseline methods}
	We use MMoE as a model of experimental comparison, which is a popular multi task learning model in industry.\\
	\subsubsection{Model hyper-parameter setup}
	In this experiment, we set the hyper-parameters for DCRNN: the input subsequence length of RNNs is set to 5, the subsequence interval of each task input is 2, and the overall sequence length can be calculated as 7. We have also conducted out unidirectional / bidirectional RNN, LSTM / GRU, is adaptive or not cross experiments and the standard control model MMoE.\\
	\subsubsection{Evaluation}
	The offline evaluation is AUC of click task and valid play task, and the online evaluation is click through rate (CTR) and valid playback rate (VPR).\\
	\subsubsection{Results}
	We compared offline indicators and online indicators respectively.\\
	\begin{table}[ht]
		\caption{Offline evaluations of Xiaomi Radio}\smallskip
		\centering
		\resizebox{\columnwidth}{!}{
		\smallskip\begin{tabular}{ccc}
			\hline  
			Model & AUC/Click & AUC/Valid play \\\hline
			MMoE & 0.6922 & 0.6721\\\hline
			DCRNN+GRU+Ada & 0.6986 & 0.6938 \\
			DCRNN+BiGRU+Ada & 0.6961 & 0.6883 \\
			DCRNN+LSTM+Ada & 0.6972 & \textbf{0.6949} \\
			DCRNN+BiLSTM+Ada & \textbf{0.6995} & 0.6804 \\\hline  
		\end{tabular}}
		\label{table1}
	\end{table}
	\begin{table}[ht]
		\caption{Influence of feature adaptive learning on Model}\smallskip
		\centering
		\resizebox{\columnwidth}{!}{
		\smallskip\begin{tabular}{ccc}
			\hline  
			Model & AUC/Click & AUC/Valid play \\\hline
			DCRNN+BiGRU & 0.6952 & 0.6748 \\
			DCRNN+BiGRU+Ada & 0.6961 & 0.6883 \\\hline
			DCRNN+BiLSTM & 0.6961 &  0.6764 \\
			DCRNN+BiLSTM+Ada & 0.6995 & 0.6804 \\\hline
			DCRNN+GRU & 0.6943 & 0.6977 \\
			DCRNN+GRU+Ada & 0.6986 & 0.6938 \\\hline
			DCRNN+LSTM & 0.6940 & 0.6827 \\
			DCRNN+LSTM+Ada & 0.6972 & 0.6949 \\\hline  
		\end{tabular}}
		\label{table2}
	\end{table}
	\paragraph{Offline Evaluation}
	From Table \ref{table1}, we can see that the DCRNN models with different settings exceed the baseline model MMoE in both indicators. However, in Table \ref{table2}, we can also see that: 1. The models with the highest performance on both indicators are not uniform; 2.From the overall experimental point of view, the application of Ada improves the performance of the model in offline indicators, but it does not improve the effective playback under DCRNN+GRU. Based on the above experiments, we finally select DCRNN+BiLSTM+Ada which performs well offline and compare it with MMoE when deployed online.\\

	\begin{table}[ht]
		\caption{Online performance of models}\smallskip
		\centering
		\resizebox{\columnwidth}{!}{
		\smallskip\begin{tabular}{ccccc}
			\hline  
			Model & CTR & PBR & \#Param\\
			\hline
			MMoE & 11.43\% &  10.17\% & 5786388 \\
			DCRNN+BiLSTM+Ada & 12.17\% &  10.93\% & 1424452 \\
			\hline
			Relative improvement & 6.47\% & 7.47\% & / \\
			\hline  
		\end{tabular}}
		\label{table3}
	\end{table}
	\paragraph{Online Evaluation}
	From Table \ref{table3}, it can be seen that DCRNN+BiLSTM+Ada perform much better online than MMoE model in terms of number of parameters, which is conducive to the deployment of online model services.\\
	\subsection{Ali-CCP Data}
	In this section, we have done some experiments on "Alibaba Click and Conversion Prediction (Ali-CCP)" to verify the effectiveness of DCRNN.\\
	\subsubsection{Task \& Data Introduction}
	Ali-CCP is a dataset gathered from real-world traffic logs of the recommender system in Taobao, and the dataset contains click and conversion data. Therefore, we make click and conversion as training targets. In this dataset, there are approximately 42 million data in the training set and test set. We sampled the two datasets without replacement, and extracted 40 million data respectively as the experimental sample set. In addition, it should be pointed out that we did not perform too much feature engineering processing on the original data, made directly "feature\_id" as the input of the model, and deleted the "feature\_value".\\
	\subsubsection{Setup}
	In this experiment, the hyper-parameters of DCRNN are set as follows: the input subsequence length of RNNs is set to 3, the subsequence interval of each task is 1, and the overall sequence length is 4. Based on the experience of Xiaomi radio experiment, we choose DCRNN+BiLSTM / DCRNN+BiLSTM+Ada. Similarly, we take MMOE as the baseline model. For the MMoE model, we set up 8 experts, each of which has 2 hidden layers, and the hidden units of each expert is 128, 64. According to the statistics of Ali-CCP dataset, exposure sample: click sample $\approx$ 24:1, exposure sample: transformation sample $\approx$ 4584:1. Therefore, in order to balance the impact of sample set imbalance on model training, we use the “tf.nn.weighted\_cross\_entropy\_with\_logits\footnote{https://www.tensorflow.org/api\_docs/python/tf/nn/weighted\_cr\\oss\_entropy\_with\_logits}” as part of the loss function. In addition, we have some other settings for the experiment: epoch=3, batch=1024, embedding\_siz=32, learning\_rate=0.0001.\\
	\subsubsection{Results}
	\begin{table}[ht]
		\caption{Model performance of Ali-CCP dataset}\smallskip
		\centering
		\resizebox{\columnwidth}{!}{
		\smallskip\begin{tabular}{ccccc}
			\hline \multirow{2}{*} { Model } & \multicolumn{2}{c} { AUC/Click } & \multicolumn{2}{c} { AUC/Conversion } \\
			\cline { 2 - 5 } & Best & Mean & Best & Mean \\
			\hline MMOE & 0.5991 & 0.5962 & 0.6545 & 0.6530 \\
			\hline DCRNN+BiLSTM & 0.6068 & \textbf{0.6067} & \textbf{0.6645} & \textbf{0.6635} \\
			\hline DCRNN+BiLSTM+Ada & \textbf{0.6076} & 0.6064 & 0.6641 & 0.6634 \\
			\hline
		\end{tabular}}
		\label{table4}
	\end{table}

	From Table \ref{table4}, the DCRNN is better than the MMOE, but in this experiment, the adaptive function does not seem to improve the fitting ability of the DCRNN, because we only use the category features as the feature input of this experiment, and do not use the statistical numerical features.\\
	
	\section{Conclusion}
	In this paper, we propose a novel multi task learning model DCRNN. DCRNN not only uses RNN to cross-process features and implicitly model multi-task sequence correlation, but it can also learn original features adaptively. In addition, instead of using the soft and hard parameter sharing architecture, we also proposed the concept of "partial parameter sharing". In experiments on Xiaomi Radio’s recommendation and “Ali-CCP” dataset, our model outperforms the baseline model MMoE. This proves the effectiveness of DCRNN in model design. In addition, the number of parameters of the DCRNN model is much smaller than that of the MMoE, which is conducive to the online service deployment of the model in real business. \\

	\bibliography{ref}
        \balance
	\bibliographystyle{aaai21}
	
\end{document}